%% file: main.tex
\documentclass[11pt]{article}
\usepackage[margin=1in]{geometry}
\usepackage{amsthm}
\usepackage{hyperref}
\usepackage{makecell} % For adto which it corresponds.e's column name
\usepackage{subcaption} % for subtables
\usepackage{natbib}
\usepackage{graphicx}
\usepackage{amsmath}
\usepackage{amssymb}
\usepackage{booktabs}
\usepackage{diagbox} 

\usepackage{authblk}

\usepackage{makecell}
\usepackage{colortbl}
\usepackage{xcolor}

\definecolor{ao}{rgb}{0.0, 0.5, 0.0}
\definecolor{vz}{rgb}{0.0, 0.0, 0}
\def\Dzero{D^{(0)}([0,1]^{d})}

\def\Dkzero{D^{(m)}([0,1]^{d})}
\def\Done{D^{(1)}([0,1]^{d})}

\def\onetod{\{1, \ldots, d\}}

\def\ibeta{\beta^{\textit{init}}}

\def\halbeta{\beta^{\textit{hal}}}

\def\rbeta{\beta^{\textit{relax}}}
\def\betaLone{\|\beta\|_1}

\def\tpsi{\psi^{\textit{target}}}
\def\rpsi{\psi^{\textit{relax}}}

\def\dto{\xrightarrow{d}}

\def\EICxbetank{D_{P_n, \phi_k, \tx}}

\def\textlu{\textit{local-u}}
\def\textgu{\textit{global-u}}

\def\hatpsi{\hat{\psi}}
\def\kcv{k_{cv}}
\def\tx{\tilde{x}}
\def\sigmahat{\hat{\sigma}}

\def\halsigmahat{\hat{\sigma}^{\textit{hal}}}
\def\relaxsigmahat{\hat{\sigma}^{\textit{relax}}}

\def\R{\mathbb{R}}

\def\bQ{\bar{Q}}

\def\g0t{\mathbf{g_0^t}}
\def\g1{\mathbf{g_1}}

\def\dto{\overset{d}{\to}}

\newcommand{\argmin}{\mathop{\mathrm{argmin}}}
\newcommand{\argmax}{\mathop{\mathrm{argmax}}}

\begin{document}

%title
\title{Constructing Confidence Intervals for Infinite-Dimensional Functional Parameters by Highly Adaptive Lasso}
\author[1]{Wenxin Zhang}
\author[1]{Junming Shi}
\author[1]{Alan Hubbard}
\author[1]{Mark van der Laan}
\affil[1]{Division of Biostatistics, University of California, Berkeley}
\maketitle

\bibliographystyle{plainnat}
\newtheorem{assumption}{Assumption}
% \newtheorem{theorem}{Theorem}
% \newtheorem{proposition}{Proposition}
% \newtheorem{result}{Result}
% \newtheorem{lemma}{Lemma}
% \newtheorem{corollary}{Corollary}

% Framework
% \begin{enumerate}
% \end{enumerate}

% \section{Summary}
\vspace{-20pt}

\begin{abstract}
Estimating the conditional mean function is a central task in statistical learning. In this paper, we consider estimation and inference for a nonparametric class of real-valued càdlàg functions with bounded sectional variation \citep{gill1995inefficient}, using the Highly Adaptive Lasso (HAL) \citep{van2015generally,benkeser2016highly,van2023higher}, a well-established empirical risk minimizer over linear combinations of tensor products of zero or higher order spline basis functions under an $L_1$-norm constraint.
Building on theoretical advances in pointwise asymptotic normality and uniform convergence rates for higher-order spline HAL estimators \citep{van2023higher}, this work focuses on constructing robust confidence intervals for HAL-based estimators of conditional means.
First, we propose a targeted HAL with a debiasing step to remove the regularization bias of the targeted conditional mean and also consider a relaxed HAL estimator to reduce such bias within the working model.
Second, we propose both global and local undersmoothing strategies to adaptively enlarge the working model and further reduce bias relative to variance. 
Third, we combine these estimation strategies with delta-method-based variance estimators to construct confidence intervals for the conditional mean.
Through extensive simulation studies, we evaluate different combinations of our estimation procedures, model selection strategies, and confidence-interval constructions. The results show that our proposed approaches substantially reduce bias relative to variance and yield confidence intervals with coverage rates close to nominal levels across different scenarios; we also provide practical guidance for choosing among these strategies for different objectives.
Finally, we demonstrate the general applicability of our framework by estimating conditional average treatment effect (CATE) functions, highlighting how HAL-based inference methods extend to other infinite-dimensional, non-pathwise-differentiable parameters.

\end{abstract}

\newpage

\section{Introduction}
Highly Adaptive Lasso (HAL) \citep{van2015generally,benkeser2016highly} is a flexible nonparametric regression estimator that minimizes empirical risk over the class of càdlàg (right continuous with left limits) functions with bounded sectional variation norms \citep{gill1995inefficient}. Rather than relying on local smoothness assumptions, HAL imposes a global smoothness constraint and achieves a fast $n^{-1/3}$ convergence rate (up to a logarithmic factor) with zero-order splines \citep{bibaut2019fast}. Recent work by \citet{van2023higher} studies higher-order smoothness classes of càdlàg functions and shows that higher-order spline HAL has asymptotic normality and uniform convergence rate $n^{-\frac{m+1}{2m+3}}$ up to a logarithmic factor ($m = 1,2,\cdots$).

Building upon these advances, this work focuses on developing valid statistical inference procedures for higher-order spline HAL. 
Unlike the goal of risk minimization, providing statistical inference requires controlling bias relative to variance, which necessitates strategies beyond the standard cross-validated HAL fit that primarily targets mean squared error. To address the unique challenges of providing HAL-based inference for estimating conditional means, we investigate several strategies designed to reduce bias in pointwise estimator. 
First, we propose global and local undersmoothing strategies to adaptively select larger working models that target the need to provide inference. Second, we introduce Targeted HAL, which aims to reduce the regularization bias present in standard HAL estimators, particularly for conditional mean estimation. Finally, together with these different types of HAL estimators, we provide Wald-type confidence intervals using the delta method based on the HAL working models.
We evaluate these point estimators, model selection strategies, and delta-method-based confidence intervals through extensive simulation studies, and provide practical recommendations tailored to different estimation goals.

Our approach is generally applicable to constructing HAL-based confidence intervals for other infinite-dimensional, non-pathwise-differentiable causal parameters, such as the conditional average treatment effect (CATE) with continuous covariates. We demonstrate how to estimate and conduct inference on CATE using our proposed HAL-based methods. Although CATE itself is not pathwise differentiable, it can be estimated by regressing doubly robust pseudo-outcomes \citep{van_der_laan_targeted_2006, van2014targeted, kennedy2023towards} on baseline covariates using flexible learners such as HAL. Thus, the central question becomes how to provide valid inference for the conditional mean function estimated by HAL, a question that is the primary focus of this study. 
 
The remainder of this paper is organized as follows. Section \ref{sec:intro_hal} introduces the definitions of càdlàg functions and HAL estimators with zero-order and first-order splines, along with their theoretical foundations for higher-order HALs.
Section \ref{sec:pointest} discusses the regular HAL estimator and introduces two variations: targeted HAL and relaxed HAL, which are designed to minimize bias of the conditional mean according to the working model. This section also describes two undersmoothing methods specifically designed to select enlarged working models to further reduce bias relative to variance.
Section \ref{section:point_est_sims} presents a series of simulation studies that assess the effectiveness of various combinations of HAL estimators and working model selectors. 
Section \ref{sec:varest} outlines methods for constructing confidence intervals using the delta method based on the HAL working model.
The evaluation of the performance of different confidence intervals, constructed from these variance estimators and combined with various point estimators and model selectors, is discussed in Section \ref{sec:ci}.
Section \ref{sec:CATE} demonstrates the application of our proposed methods to provide HAL-based estimation and inference for CATE. We summarize our findings and conclude in Section \ref{sec:discussion}.

\section{Defining Functional Estimation Problem and HAL-MLEs}
\label{sec:intro_hal}

\subsection{Zero-Order HAL}
Let $\Dzero$ be a class of $d$-variate càdlàg functions defined on a unit cube $[0,1]^d$ with a bounded sectional variation norm \citep{gill1995inefficient}. 
For any function $Q \in \Dzero$,
define a function $Q_s: (0_s,1_s] \to \R$ by $Q_s(u(s)):=Q(u(s),0(-s))$ for every non-empty $s \subset \onetod$, which sets the coordinates in the complement of subset $s$ equal to zero.
The sectional variation norm of $Q$ is defined by 
$\|Q\|_v^* := \int_{[0,1]^d}|d Q(u)|=|Q(0)|+\sum_{s \subset [d], s \ne \emptyset} \int_{(0(s), 1(s)]}|Q(d u(s), 0(-s))|
$.
If $\|Q\|_v^*$ is finite, then any càdlàg function $Q \in \Dzero$ can be expressed as follows:
\[
Q(x) = \int_{[0, x]} d Q(u) := Q(0)+\sum_{\substack{s\subset[d]\\ s\neq\emptyset}} \int_{(0(s), 1(s)]} \phi_u^0(x(s)) Q(d u(s), 0(-s)),
\]
where $\phi_u^0$ is a zero-order basis function indexed by a knot point $u$, defined by $\phi_u^0(x) := I(x \geq u)=\prod_{j=1}^d I\left(x_j \geq u_j \right)$, a tensor product of zero-order splines.

With this representation, one can approximate $Q$ by a discrete measure with a finite number of support points (say $l$)  \citep{van2015generally, benkeser2016highly}. For each subset $s \subset \{1,\cdots,d\}$, define $Q_{l,s}$ as a discrete approximation of $Q_s$ with support points $(u_{s,j}:j)$. Then 
$Q(x)$ can be approximated by
$
Q_l(x) := Q(0) + \sum_{s \subset [d], s \ne \emptyset} \sum_{j} \phi_{u_{s,j}}^0(x) dQ_{l,s,j}.
$
The sectional variation norm of $Q_l$ is given by $\|Q_l\|_v^* :=  |Q(0)| + \sum_{s \subset [d], s \ne \emptyset} \sum_{j} |dQ_{l,s,j}|$.

Suppose that one observes $n$ i.i.d. samples $(x_i,y_i)$ ($i = 1,\cdots,n$) with $x_i = (x_{i,1},\cdots,x_{i,d}) \in [0,1]^d$, $y_i \in \R$. Each sample $i$ naturally provides a support point $x_i(s)$ on the subset $s$. Then, one can approximate $Q$ by $Q_{\beta}$, which is defined by 
\[
Q_{\beta}(x) := \beta_0 +  \sum_{s \subset [d], s \ne \emptyset} \sum_{i=1}^n \beta_{s,i}\phi_{x_i(s)}^0(x(s)) = \beta_0 +  \sum_{s \subset [d], s \ne \emptyset} \sum_{i=1}^n \beta_{s,i} I(x(s) \geq x_i(s)).
\]

Define $P f:= \int f(o) dP(o)$ for a function $f$ with a data generating distribution $P$ that generates data $O \sim P$, and denote $P_n f:= \int f(o) dP_n(o)$ under empirical distribution $P_n$.
A zero-order HAL-MLE can be obtained by minimizing the empirical risk over the class
\[
\mathcal{Q}_{n,C} := \left\{ Q_{\beta}:  |\beta_0| +  \sum_{\substack{s\subset[d]\\ s\neq\emptyset}}  \sum_{i=1}^n |\beta_{s,i}| < C \right\},
\]
for some constant $C$ and a chosen loss function $L(Q): \mathcal{O} \to \R$ corresponding to $Q \in \Dzero$. That is, 
\[
Q_n := \argmin_{\beta: \, |\beta_0|+ \sum_{s \subset [d], s \ne \emptyset}\sum_{i=1}^n |\beta_{s,i}| < C_n^u} P_n L(Q_\beta),
\]
where $C_n^u$ is a data-adaptive cross-validation or undersmoothed selector of bounded sectional variation norm for the working model with basis functions of $\phi_{x_i(s)}^0$'s.
\citet{bibaut2019fast} shows that zero-order HAL-MLE attains an $L_2$ rate of convergence of $n^{-\frac{1}{3}}$ up to a logarithmic factor.

\subsection{First-Order and Higher-Order HALs}
\citet{van2023higher} further defines a $m$-th order smoothness class $\Dkzero$ $(m = 1, 2, \cdots)$ on $[0,1]^d$, which includes functions whose $m$-th-order Lebesgue-Radon-Nikodym derivatives are themselves real-valued càdlàg functions with bounded sectional variation norm.
Like the zero-order HAL MLE, one can construct a higher order HAL-MLE that uses $m$-th order splines to estimate the target function in $\Dkzero$. For the first order ($m=1$), one can approximate $Q^{(1)} \in \Done$ by $Q^{(1)}_{\beta}$, which can be written by 
\begin{equation}
    Q^{(1)}_{\beta}(x) := \beta_0 +  \sum_{s \subset \{1,\cdots,d\}, s \neq \emptyset} \sum_{i=1}^n \beta_{s,i}\phi_{x_i(s)}^1(x(s)), 
    \label{eq:hal_mle}
\end{equation}
where $\phi^1_u(x) = (x - u)I(x \ge u)$.
The first-order HAL-MLE is given by 
\[
Q_n^{(1)} := \argmin_{\beta: \, |\beta_0|+ \sum_{s \subset \{1,\cdots,d\}, s \neq \emptyset} \sum_{i=1}^n |\beta_{s,i}| < C_n^u} P_n L(Q_\beta^{(1)}).
\]
One can also consider the second-order spline function, which is $\phi^2_u(x) = \frac{1}{2}(x - u)^2 I(x \ge u)$ or higher-order spline basis functions to fit higher-order HAL-MLEs.
Moreover, \citet{van2023higher} establishes uniform covering number and entropy bounds for these higher-order HAL estimators, and shows that the 
$m$-th order spline HAL-MLE achieves asymptotic normality and a uniform convergence rate of $n^{-\frac{m+1}{2m+3}}$ up to logarithmic factors ($m = 1,2,\cdots$).
These theoretical results provide the foundation for the construction of confidence intervals for conditional means based on $m$-th order HAL working models, as investigated in this paper.

\section{Pointwise Estimation}
\label{sec:pointest}

\subsection{Estimation and Inference Using HAL Working Model}
\subsubsection{Regular HAL}
Suppose we have obtained $K$ candidate HAL working models by fitting HAL under a series of strictly increasing $L_1$-norm constraints $M_1 < \cdots <M_K$. 
For each bound $M_k$, we retain only the $s_k$ basis functions whose estimated coefficients are non-zero. We enumerate these basis functions as $\phi_{n,k} = (\phi_{n,k,1},\cdots,\phi_{n,k,s_k})$ and define the associated vector-valued map $\phi_{n,k}(x):=\big(\phi_{n,k,1}(x),\ldots,\phi_{n,k,s_k}(x)\big)^\top \in \R^{s_k}$ for $x \in [0,1]^d$.
Let the corresponding coefficient vector be $\halbeta_{n,k} := (\halbeta_{n,k,1},\cdots,\halbeta_{n,k,s_k})^{\top} \in \R^{s_k}$. The resulting HAL fit can be written as $Q_{n,k}(x)=\phi_{n,k}(x)^{\top}\halbeta_{n,k}$.
We further define $Q_{0,k}$ as the $L_2(P_0)$-projection of the true conditional mean function $Q_0$ onto the linear span of $\phi_k$. Specifically, $Q_{0,k}(x) = \phi_k(x)^{\top}\beta_{0,k}$, where 
$
\beta_{0,k} : = \argmin_{\beta \in \R^{s_k}} P_0 (Y - \phi_k(X)^{\top} \beta)^2.
$

For any test point $\tx = (\tx_1,\cdots,\tx_d) \in [0,1]^d$, we use
$
\phi_k(\tx) = (\phi_{k,1}(\tx),\cdots,\phi_{k,s_k}(\tx))^{\top} \in \R^{s_k}$
to denote a transformation of $\tx$ by basis functions in $\phi_k$. 
For a given point $\tx$, we use $\psi(\tx) :=  Q_{0}(\tx) = E(Y|X=\tx)$ to denote the true conditional mean of $Y$  given $\tx$.
Its projection onto the $k$-th HAL working model is
$
\psi_k(\tilde{x}):=Q_{0,k}(\tilde{x})=\phi_k(\tilde{x})^\top\beta_{0,k}$.

We note that $\halbeta_{n,k}$ in regular HAL is subject to the $L_1$-norm constraint $\betaLone \leq M_k$. 
This may result in a choice of less complex model compared to the true model, and the linear combination of basis functions in $\phi_k$ may not well approximate $\psi_k$, resulting in a projection bias between the true target estimand $\psi(\tx)$ and $\psi_k(\tx) = \phi_k(\tx)^{\top} \beta_{0,k}$.
On the other hand, even when the linear combination of basis functions in $\phi_k$ is rich enough to approximate the true function, the regular HAL estimator $\phi_k(\tx)^{\top} \beta_{n,k}$ might be subject to regularization bias from $\beta_{0,k}$ within the working model represented by $\phi_k$.
That is because the score equation $P_n \EICxbetank(\halbeta_{n,k})$, which will be introduced in the next subsection,
may not be fully solved to zero due to the $L_1$ constraint $\betaLone \le M_k$.

We explore several solutions to mitigate two potential sources of bias. 
In Section \ref{relaxed-HAL} and \ref{target-HAL}, we introduce two additional HAL estimators to address the regularization bias. 
In Section \ref{undersmooth}, we present two undersmoothing techniques designed to counteract potential projection bias.
For notational convenience, let \(\psi_{\tx}(\beta)\) denote the point estimate a test point \(\tx\) obtained from a non-zero coefficient vector \(\beta\) and the associated basis vector\(\phi\). In particular, for the \(k\)-th working model with basis \(\phi_k\) and non-zero coefficients $\beta_k$, we write \(\psi_{\tx}(\beta_k):=\phi_k(\tx)^{\top} \beta_k\).

\subsubsection{Relaxed HAL}
\label{relaxed-HAL}
Suppose that the true distribution of $P_0$ follows a working model which is linearly represented by the basis functions $\phi_{k}$, then $\beta_{0,k}$ is the solution of a score equation 
$P_0 \phi_k(X) (Y- \phi^{\top}_k(X)\beta) = 0$.
The efficient influence function of $\beta_{0,k}$ at $P_0$ is 
\[D_{P_0, \phi_k}(\beta_{0,k})(O) := P_0\left(\phi_k(X) \phi_k(X)^{\top}\right)^{-1} \phi_k(X) (Y - \phi_k(X)^{\top}\beta_{0,k} ).\]

We define $\rbeta_{n,k} := argmin_{\beta \in \R^{s_k}} \sum_{i = 1}^{n}(Y_i - \phi^{\top}_k(X_i)\beta)^2$, using the superscript ``\textit{relax}'' to imply that $\rbeta_{n,k}$ is refitted under the working model without $L_1$ restriction $\betaLone \le M_k$, thus it completely solves the score equation \citep{van2023adaptive}.

We use $\Phi_{n,k}:= (\phi_k(X_1), \cdots, \phi_k(X_n))^{\top}$ to denote a $n \times s_k$ matrix obtained by applying $s_k$ basis functions within $\phi_k$ to the design matrix  $(X_1,\cdots,X_n)^{\top}$ in the training data.
By applying the delta method, we obtain the efficient influence function of conditional mean at the point $\tx$ under $P_0$ as follows:
\[
D_{P_0, \phi_k,\tx}(\beta_{0,k})(O) := \phi_k(\tx)^{\top} P_0 \left(\phi_k(X) \phi_k(X)^{\top}\right)^{-1} \phi_k(X) (Y - \phi_k(X)^{\top}\beta_{0,k}).
\]

Thus we have
\[
\sqrt{n}(\phi_k(\tx)^{\top} \rbeta_{n,k} - \phi_k(\tx)^{\top} \beta_{0,k}) \dto N(0, V_{0,k, \tx})
\]
under some regular assumptions in \citet{van2000asymptotic}, 
where $V_{0,k,\tx} := P \left(D_{P_0, \phi_k,\tx}(\beta_{0,k})\right)^2$ and can be estimated by
$V_{n,k,\tx}(\rbeta_{n,k}) := \frac{1}{n} \sum_{i=1}^n \left(D_{P_n, \phi_k,\tx}(\rbeta_{n,k})(O_i)^2\right)$. 
The estimate of $E[Y|X=\tx]$ based on the relaxed HAL is then given by $\rpsi_{\tx}(\rbeta_{n,k}) := \phi_k(\tx)^\top \rbeta_{n,k}$.
Unlike the regular HAL-MLE $\phi_k(\tx)^{\top}\halbeta_{n,k}$ which is constrained by $\halbeta_{n,k} \leq M_k$, the relaxed HAL is not subject to that restriction.

\subsubsection{Targeted HAL}
\label{target-HAL}
We consider the case where one obtains an initial estimate of $\ibeta_{n,k}$ for $\beta_{0,k}$ in each working model represented by $\phi_k$ (e.g., the working model selected by cross-validation selector) and consequently obtain an initial estimate $\psi_{\tx}(\ibeta_{n,k}) := \phi_k(\tx)^{\top} \ibeta_{n,k}$ of $\psi_{\tx}(\beta)$. A one-step debiasing procedure \textit{targeted} for reducing bias in \(\psi_{\tx}(\beta)\) can be applied by updating $\psi_{\tx}(\ibeta_{n,k})$ as follows:
\[
\tpsi_{\tx}(\ibeta_{n,k}):= \psi_{\tx}(\ibeta_{n,k}) + P_n D_{P_n, \phi_k,\tx}(\ibeta_{n,k}).
\] 
If we treat $\phi_k$ as a fixed set of basis functions, then under some regular assumptions we have that
\[
\sqrt{n}\left( \tpsi_{\tx}(\ibeta_{n,k}) - \phi_k(\tx)^{\top} \beta_{0,k} \right) \dto N(0, V_{0,k, \tx}),
\]
where $V_{0,k,\tx}$ can be estimated by 
$V_{n,k,\tx}(\ibeta_{n,k}) := \frac{1}{n} \sum_{i=1}^n \left(D_{P_n, \phi_k,\tx}(\ibeta_{n,k})(O_i)^2\right)$. 
In practice, running regular HAL with cross-validation procedure naturally returns $K$ working models and $K$ sets of coefficients
$\halbeta_{n,k}$ as initial estimates of $\beta_{0,k}$.
Therefore, we can directly obtain a point estimator $\tpsi_{\tx}(\halbeta_{n,k})$ using this one-step debiasing procedure:
$
\tpsi_{\tx}(\halbeta_{n,k}):= \psi_{\tx}(\halbeta_{n,k}) + P_n D_{P_n, \phi_k,\tx}(\halbeta_{n,k}).
$
This debiasing step removes the regularization bias of $\halbeta_{n,k}$ within the working model, which is more convenient than rerunning a relaxed HAL because it does not need to refit the model on the training data.

\subsection{Cross-validation and Undersmoothing Selectors for Working Models} 
\label{undersmooth}
\subsubsection{Cross-validation Selector for HAL}
Running regular HAL returns multiple models that differ in their sectional variation norms. This necessitates a model-selection strategy that serves for various objectives for providing robust inference. The regular model selector of HAL chooses one of the candidate working models such that its regular HAL-MLE minimizes the cross-validated mean squared error (CV-MSE), which balances bias and variance. We refer to this selection method as ``cross-validation selector'' of the working model.
However, when constructing confidence intervals, it is crucial to prioritize reducing bias relative to variance. Therefore, the typical CV-MSE minimization approach may not be ideal, as it does not sufficiently prioritize reducing bias in the context of confidence interval construction.
Moreover, when our target is to provide a confidence interval for the conditional mean of a user-specific test point, the model returned by the cross-validation selector was not targeted for reducing bias for that specific test point. 
When the functional form of the working model is overly simplistic to estimate the target, it is possible that projection bias $Q_{0,k}(\tx) - Q_0(\tx)$ arises for some test points $\tx$.
On the other hand, even with a sufficiently large working model, we may still encounter regularization bias from the $L_1$ restriction.

To address these issues, we propose two undersmoothing methods for HAL-based inference for general conditional mean functions.
Undersmoothing methods intentionally select working models subject to an $L_1$-norm bound that is larger than that selected by cross-validation. This helps to mitigate the risks of both projection bias and regularization bias, by admitting more basis functions to construct the working model and reducing the impact of penalty-driven shrinkage on coefficient estimates. 

\subsubsection{Locally Undersmoothed HAL}
Recall that we have $K$ candidate working models represented by $K$ sets of basis functions $\phi_{1}, \cdots, \phi_{K}$. Instead of the cross-validation-selected working model used in the regular HAL, we propose an undersmoothing strategy to incrementally move towards a candidate working model with larger sectional variation norm.
This deliberate model expansion continues until a measure of the ratio of bias versus standard error of the estimator is sufficiently small, which is a critical condition for constructing a confidence interval with desirable coverage properties. 
Specifically, for a single evaluation point $\tx$, this undersmoothing method selects the first model that satisfies the following criteria:
\[
k_{\textlu} := \min_k \left\{k \in \{\kcv,\cdots,K\}: 
\frac{|P_n D_{P_n, \phi_k,\tx}(\halbeta_{n,k})|}{\sqrt{P_n \left(D_{P_n, \phi_{\kcv},\tx}(\halbeta_{n,\kcv})\right)^2/n}} \le \frac{1}{log(n)}\right\},
\]
which we call ``local undersmoothing''.
The locally undersmoothed HAL-MLE of $E(Y|X=\tx)$ is $\psi_{\tx}(\halbeta_{n, k_{\textlu}}) := \phi_{k_{\textlu}}(\tx)^\top \halbeta_{n,k_{\textlu}}$.

The spirit of this undersmoothing method is to expand the working model until the bias versus standard error is sufficiently small for a good coverage rate of our confidence intervals \footnote{We use the threshold $\frac{1}{\log(n)}$ as an undersmoothing criterion following \citet{van2019causal}, where an undersmoothed version of HAL is applied to obtain an efficient plug-in estimator for pathwise-differentiable parameters.}. Both bias and standard error are defined by the efficient influence function of the conditional mean under the working model. To facilitate a consistent and objective evaluation of the performance across various undersmoothed working models $\phi_k$, particularly concerning their capacity for bias reduction, we employ the standard error $\frac{1}{\sqrt{n}}\sqrt{P_n\left( D_{P_n, \phi_{\kcv},\tx}(\halbeta_{n,\kcv})\right)^2}$ as the denominator.
This choice is strategic: by maintaining the standard error as a fixed reference point — specifically that derived from applying the delta method to the coefficients under the working model selected by the cross-validation procedure - we ensure that all comparisons are grounded on a common baseline.

\subsubsection{Globally Undersmoothed HAL}
Furthermore, consider a scenario with 
$J$ distinct test points $(\tx_1,\cdots,\tx_J)$.
When constructing confidence intervals for multiple test points, one might pursue a single undersmoothed model instead of separate models for each point.
To achieve this, we propose a global undersmoothing approach aimed at selecting a working model $Q_n(x)$, denoted as $\phi_{k_{\textgu}}$, such that,

\[
k_{\textgu} := \argmax_{k \in \{\kcv,\cdots,K\}} \frac{1}{J} 
\sum_{j = 1}^{J} 
I\left(
\frac{|P_n D_{P_n, \phi_k,\tx_j}(\halbeta_{n,k})|}{\sqrt{\frac{1}{n}P_n \left(D_{P_n, \phi_{\kcv},\tx_j}(\halbeta_{n,\kcv})\right)^2}} \le \frac{1}{log(n)}
\right).
\]
With a globally undersmoothed working model represented by basis functions indexed by $k_{\textgu}$,
the globally undersmoothed HAL-MLE for every $E(Y|X=\tx_j)$ is  $\phi_{k_{\textgu}}(\tx_j)^\top \halbeta_{n,k_{\textgu}}$.

Since the global undersmoothing aims to find only one working model across all test points, its performance is not guaranteed to always align with the 
working models identified by local undersmoothing tailored to each point, particularly when these points exhibit heterogeneity in how well their conditional expectations can be approximated by working models with different complexity.

We also note that, in practice, undersmoothing alone may not fully eliminate regularization bias in finite samples. 
For example, when implementing undersmoothing, one might want to cap the $L_1$-norm of the working models to avoid including too many basis functions, which can still induce regularization bias. 
To address this, we complement the undersmoothing approach with targeted HAL-MLEs for estimating $E(Y|X=\tx)$: 
\begin{align*}
\tpsi_{\tx}(\halbeta_{n,k_{\textlu}})&:= \phi_{k_{\textlu}}(\tx)^{\top} \halbeta_{n,k_{\textlu}} + P_n D_{P_n, \phi_{k_{\textlu}},\tx}(\halbeta_{n,k_{\textlu}});\\
\tpsi_{\tx} (\halbeta_{n,k_{\textgu}}) &:= \phi_{k_{\textgu}}(\tx)^{\top} \halbeta_{n,k_{\textgu}} + P_n D_{P_n, \phi_{k_{\textgu}},\tx}(\halbeta_{n,k_{\textgu}}).
\end{align*}

\section{Simulation Study of Point Estimator Performance}
\label{section:point_est_sims}
% Possibly select some plots to show in the main paper instead of all of them?
In this section, we implement a series of numerical studies to evaluate the finite-sample performance of various point estimators across different scenarios. 
We evaluate the nine distinct HAL estimators (Table \ref{tab:hal_estimators}) based on first-order spline HALs, each derived from a combination of three working model selectors (CV, Global Undersmoothing, and Local Undersmoothing) and three pointwise estimation methods (regular HAL, targeted HAL and relaxed HAL) based on the selected working models as we discussed above.

\begin{table}[htbp]
\centering
\small
\caption{
Nine HAL point estimators formed by combining three working model selectors (rows) and three estimation methods (columns). Each label (e.g., \texttt{cv.targeted}) denotes a unique estimator configuration.
}
\label{tab:hal_estimators}
\begin{tabular}{|c|c|c|c|}
\hline
\diagbox[width=15em]{\textbf{Model Selector}}{\textbf{Estimation}}
& Regular HAL
& Targeted HAL 
& Relaxed HAL \\
\hline
Cross-Validation & cv.regular & cv.targeted & cv.relax \\
\hline
Global Undersmoothing & global-u.regular & global-u.targeted & global-u.relax \\
\hline
Local Undersmoothing & local-u.regular & local-u.targeted & local-u.relax \\
\hline
\end{tabular}
\end{table}

\subsection{Simulation Setup}
\label{section:simulation_setup}
In simulations, each estimator is trained using sample sizes of 250, 500, 1000, and 2000, respectively. We then measure their performance over 20 test points, which are generated from the same distributions.
We examine three types of data generating distributions and, for each, consider real-valued covariates $X = (X_1,\cdots,X_d)$ with dimensions $d \in \{1, 3, 5\}$. 
The distribution of $X$ is as follows:
\[
\begin{aligned}
X_1 &\sim \operatorname{Uniform}(-4,4) &; \quad X_2 &\sim \operatorname{Uniform}(-4,4) &; \\
X_3 &\sim \operatorname{Bernoulli}(0.5) & ; \quad X_4 &\sim \operatorname{Normal}(0,1) &; \\
X_5 &\sim \operatorname{Gamma}(2,1).
\end{aligned}
\]
For each data generating distribution, $Y$ is generated by $Q_0(X)+ \epsilon$, where $Q_0$ is the true conditional mean function and $\epsilon \sim N(0,1)$.
In Scenario 1, $Q_0$ is defined as follows:
\begin{itemize}
    \item $d = 1$: $Q_0(x) = 0.05 x_1 + 0.04$;
    \item $d = 3$: $Q_0(x) = 0.07 x_1 - 0.28 x_1 x_2 + 0.05 x_2 + 0.25 x_3 x_2$;
    \item $d = 5$: $Q_0(x) = 0.1 x_1 - 0.3 x_1 x_3 + 0.25 x_2 + 0.5 x_3 x_2 - 0.5 x_4 + 0.04 x_5 x_4 - 0.1 x_5$.
\end{itemize}
In Scenario 2, $Q_0$ is given by:
\begin{itemize}
    \item $d = 1$: $Q_0(x) = 0.05 x_1 + 0.04 x_1^2$;
    \item $d = 3$: $Q_0(x) = 0.07 x_1 - 0.28 x_1^2 + 0.05 x_2 + 0.25 x_2 x_3$;
    \item $d = 5$: $Q_0(x) = 0.1 x_1 - 0.3 x_1^2 + 0.25 x_2 + 0.5 x_2 x_3 - 0.5 x_4 + 0.04 x_5^2 - 0.1 x_5$.
\end{itemize}
In Scenario 3, $Q_0$ is defined as:
\begin{itemize}
    \item $d = 1$: $Q_0(x) = \sigma(x_1)$;
    \item $d = 3$: $Q_0(x) = \sigma\left(-2 x_1 I(x_1 > -0.5) - x_3 + 2 x_2 x_3\right)$;
    \item $d = 5$: $Q_0(x) = \sigma\left(-2 x_1 I(x_1 > -0.5) - x_3 + 2 x_2 x_3 - 0.5 x_4 + x_4 x_5 - 0.25 x_5\right)$,
\end{itemize}
where $\sigma: \R \to (0,1)$ is a sigmoid function $\sigma(x) = \frac{1}{1+e^{-x}}$.

The three scenarios vary in difficulty in approximating the true function with first-order splines. 
In Scenario 1, $Q_0$ can be well approximated by first-order splines, as it only includes first-order terms. 
In Scenario 2, the inclusion of quadratic terms increases the challenge for HAL splines to approximate  $Q_0$. 
In Scenario 3, the true function's non-linear nature further complicates the approximation by first-order HAL splines.

Focusing on statistical inference, we assess each estimator's ability to reduce bias relative to variance. Additionally, we evaluate the coverage of Wald-type ``oracle confidence intervals'', which are constructed using the standard deviation of each estimator over 500 Monte Carlo runs, which we refer to as ``oracle standard error''.

\subsection{Results}
\label{section:point_est_result}

Figure \ref{fig:boxplot_bias} shows box plots of the bias of point estimators across test points for all scenarios. 
For the same type of point estimator, those based on a working model selected through local undersmoothing exhibit smaller bias compared to those chosen by global undersmoothing, which in turn show less bias than those selected via cross-validation. This pattern is consistent across all scenarios, particularly for higher dimensions.

When comparing different pointwise estimators under the same working model selector, both targeted HAL and relaxed HAL reduce bias compared to regular HAL, which is constrained by the $L_1$-norm restriction. 
Specifically, targeted HAL and relaxed HAL are particularly effective in reducing bias when $d = 3, 5$ in Scenario 1, $d \in \{3, 5\}$ in Scenario 2, and across all dimensions in Scenario 3.
The bias in these cases is further diminished with the combination of undersmoothing and targeted HAL or relaxed HAL.

\input{figures/bias}

To provide a more comprehensive understanding of the bias of all point estimators with different working model selectors, Figure \ref{fig:bias_reg_proj} presents regularization bias and projection bias of regular HAL. These are measured by the absolute value of \(\psi_{\tx}(\halbeta_{n,k}) - \psi_{\tx}(\rbeta_{n,k})\) and the absolute value of \(\psi_{\tx}(\rbeta_{n,k}) - Q_0(\tx)\).
Most scenarios show regularization bias for regular HAL in CV-selected working model.
In Scenario 1, regularization bias is larger than projection bias at $d=1$, while their scales are similar at $d=3$ and $d=5$.
In Scenario 2, for $d = 1$, the regularization bias is negligible.
For $d = 3$ and $d = 5$, the two biases are comparable.
In Scenario 3, the two biases are similar in $d=1$ and $d=3$, while the projection bias is consistently larger than the regularization bias at $d=5$.
In contrast, both undersmoothing methods effectively decrease both types of bias in all scenarios, with local undersmoothing more effective than global undersmoothing in bias reduction.
This explains why undersmoothing, combined with targeting or relaxed HAL estimators, is effective in reducing bias in most scenarios.
\input{figures/bias_reg_proj}

\input{figures/bias_se_oracle}

In Figure \ref{fig:boxplot_bias_se_oracle},
we present the ratio of bias over oracle standard error of the point estimators, which is an important metric for assessing their potential to make statistical inference.
This reduction is particularly evident across all dimensions in Scenario 1, for $d \in \{3,5\}$ in Scenario 2, and across all dimensions in Scenario 3. 
In particular, the combination of undersmoothing and targeted (or relaxed) HAL effectively reduces the ratio of bias to oracle standard error from above the recommended threshold of $\frac{1}{\log(n)}$ to below for most test points in Scenarios 1 and 2 for $d \in \{3,5\}$, and in Scenario 3 for $d \in \{1,3\}$. 
Even in the most challenging case of $d = 5$ in Scenario 3, undersmoothing and targeted (or relaxed) HAL are effective in reducing that ratio toward the threshold.

\input{figures/cov_oracle}
Finally, we study the performance of Wald-type oracle confidence intervals that are constructed using point estimators and their Monte Carlo standard deviation across 500 runs. Figure \ref{fig:cov_oracle} shows box plots of the coverage probabilities of oracle confidence intervals for all point estimators across all test points. 
For dimension $d = 1$, all combinations of point estimators and model selectors achieve the nominal coverage probability of 95\% for all test points. However, when the dimensions are $d = 3$ and $d = 5$, the oracle confidence intervals for regular HAL-MLEs fail to achieve the nominal coverage rate for some test points, particularly in Scenario 2 and Scenario 3. While undersmoothing improves coverage probabilities for some of these test points, there are still other points where the oracle confidence intervals do not perform well. 
In contrast, the coverage probabilities of oracle confidence intervals for targeted and relaxed HALs achieve the nominal 95\% coverage rate for \(d = 3\) in Scenario 2 and Scenario 3, even without undersmoothing. However, their coverage probabilities are off for some test points when $d = 5$. Undersmoothing improves their coverage rates in these challenging cases. In comparison, targeted HAL and relaxed HAL achieve the nominal oracle coverage in most scenarios without undersmoothing. In Scenario 3 and $d=5$, their oracle coverage probabilities are slightly below 95\% based on working model selected by cross-validation, while global undersmoothing improves them, and local undersmoothing further brings them closer to the nominal level, demonstrating their asymptotic normality.

\section{Variance Estimators and Confidence Intervals}
\label{sec:varest}

% \subsection{Variance Estimator based on delta method}
Recall that given a HAL working model represented by linear combinations of basis functions in $\phi_k$, one can consider applying the delta method to estimate the variance of the HAL estimate of the conditional mean of the test point $\tx$. 
Let $\halsigmahat_{n,k,\tx}(\halbeta_{n,k}) := \sqrt{\frac{1}{n}P_n \EICxbetank^2(\halbeta_{n,k})}$.
We construct 
$
[\psi_{\tx}(\halbeta_{n,k}) - z_{1-\frac{\alpha}{2}}  \halsigmahat_{n,k,\tx}, \psi_{\tx}(\halbeta_{n,k})+ z_{1-\frac{\alpha}{2}}  \halsigmahat_{n,k,\tx}],
$ and
$ 
[\tpsi_{\tx}(\halbeta_{n,k}) - z_{1-\frac{\alpha}{2}}  \halsigmahat_{n,k,\tx}(\halbeta_{n,k}), \tpsi_{\tx}(\halbeta_{n,k})+ z_{1-\frac{\alpha}{2}}  \halsigmahat_{n,k,\tx}(\halbeta_{n,k})],
$
as confidence intervals for regular HAL-MLE and targeted HAL, respectively, where $z_{1-\frac{\alpha}{2}}$ is the $1-\frac{\alpha}{2}$-quantile of standard normal distribution ($\alpha = 0.05$). 
For relaxed HAL, the confidence interval is
$
[\psi_{\tx}(\rbeta_{n,k}) - z_{1-\frac{\alpha}{2}} \relaxsigmahat_{n,k,\tx}(\rbeta_{n,k}), 
\psi_{\tx}(\rbeta_{n,k}) + z_{1-\frac{\alpha}{2}}  \relaxsigmahat_{n,k,\tx}(\rbeta_{n,k})],
$
where $\relaxsigmahat_{n,k,\tx}(\rbeta_{n,k}):= \sqrt{\frac{1}{n}P_n\EICxbetank^2(\rbeta_{n,k})}$.

We also considered a conservative confidence interval to account for possible projection error between the working model and true model. Specifically, we introduce an adaptive penalty factor to adjust the width of delta-method-based confidence interval, with the rationale being that if the within-model bias of the fitted HAL under a working model is not sufficiently small (quantified by how the score equation for the conditional mean of $\tx$ is sufficiently solved by the HAL fit indexed by its coefficient denoted by $\beta_{n,k}$), then we conservatively expect a projection error between the working model and the true model, with a similar scale. For the working model represented by the $k$-th set basis functions and the associated coefficient $\beta_{n,k}$ in the fitted HAL, we define an adaptive penalty factor \(\gamma\) as
$
\gamma_{n,k,\tx}(\beta_{n,k}) := \max(\frac{|P_n \EICxbetank(\beta_{n,k})|}{\sqrt{\frac{1}{n}P_n \EICxbetank^2(\beta_{n,k})}}, \frac{1}{log(n)}).
$
We use $\hatpsi_{n,k,\tx}$ and $\sigmahat_{n,k,\tx}$ to denote the point estimate and standard error estimator used for constructing regular confidence intervals based on the $k$-th working model introduced above.
Using the adaptive penalty factor, we define their conservative confidence interval as
$
[\hatpsi_{n,k,\tx} - z_{1-\frac{\alpha}{2}} \sigmahat_{n,k,\tx}^{consv}, \hatpsi_{n,k,\tx} + z_{1-\frac{\alpha}{2}} \sigmahat_{n,k,\tx}^{consv}],
$ where $
\sigmahat_{n,k,\tx}^{consv} = \sigmahat_{n,k,\tx} \sqrt{1+\gamma_{n,k,\tx}(\beta_{n,k})^2}.
$
Such conservative confidence intervals are more useful in regular and targeted HAL, while relaxed HAL is minimally inflated as it already solves the score equation under the working model.

\section{Simulation Study of Confidence Interval Performance}
\label{sec:ci}
In this section we evaluate variance estimators and delta-method-based confidence intervals of HAL. 
We follow the simulation setup described in Section \ref{section:simulation_setup} and use the results of oracle confidence intervals from Section \ref{section:point_est_result} as benchmarks for comparison.
\input{figures/width_delta_oracle}
In Figure \ref{fig:width_ratio}, we compare the width of confidence intervals based on the delta method with the width of the oracle confidence intervals, where the width of oracle confidence intervals is a built upon Monte Carlo standard deviation. 
These ratios compare our variance estimators' performance to the oracle variance estimator, which already achieves nominal coverage probabilities via oracle Wald-type confidence intervals, as discussed above.
The results show that the delta-method-based variance estimator tends to underestimate the true variance for targeted HAL or relaxed HAL in most scenarios with a small sample size. As the sample size increases and undersmoothing is applied, the ratio of the width of delta-method-based confidence intervals to the width of oracle confidence intervals approaches 1. 

In comparison, the delta-method-based variance estimator tends to overestimate the true variance for regular HAL-MLEs in dimensions $3$ and $5$. This may be explained by the fact that, conditional on the working model, applying the delta method to regular HAL-MLE ignores the \(L_1\)-constraint of \(\halbeta_{n,k}\), which may reduce the true variation of the estimator. Treating \(\halbeta_{n,k}\) as fitted from a regression without the \(L_1\)-constraint leads to an overestimation of the true variance conditional on the working model. 
thus providing a conservative confidence interval in finite samples.

\input{figures/cov_delta}

Finally, we study the coverage probabilities of these confidence intervals based on different combinations of model selectors, point estimators and delta-method-based variance estimators in Figure \ref{fig:cov_delta}.
As expected, confidence intervals for regular HAL-MLEs based on undersmoothed working model exhibit the highest coverage probabilities compared to those for targeted or relaxed HALs, owing to the enlarged working model with more basis approaching the true model, together with the overestimation of variance within the working model. When the sample size is small, the conservative confidence intervals with adaptive penalty factor help increase the coverage rate for regular HAL and targeted HAL. 

\section{Application to HAL-Based CATE Estimation and Inference}
\label{sec:CATE}
In this section, we demonstrate the application of the proposed methods in estimating conditional average treatment effect (CATE) and providing inference. It is implemented by regressing doubly robust pseudo-outcome \citep{van2006statistical,van2014targeted,kennedy2023towards} on covariates based on our proposed methods to make inference for conditional mean functions based on HAL, as illustrated above.
In this simulation, we observe $n$ i.i.d. data \(O=(W, A, Y) \sim P_0\) under the standard no unmeasured confounding assumption, where $W$ denotes the $d$-dimensional baseline covariates \(W\in [0,1]^d\), \(A \in \{0,1\}\) is binary treatment, and $Y \in \R$ is the outcome of interest.
Let the true propensity score function be $g_0: \{0,1\} \times [0,1]^d \to (0,1)$, where $g_0(A=a|W=w) = P_0(A = a \mid W=w)$. Define the true outcome function $\bQ_0: \{0,1\} \times \R^d \to \R$, where \(\bQ_{0}(A, W) = E_{P_0}[Y \mid A, W]\). We denote the true conditional average treatment effect (CATE) function of interest by $B_0: [0,1]^d \to \mathbb{R}$, defined by 
$
B_0(W) := E[Y \mid A=1, W] - E[Y \mid A=0, W].
$

To construct the pseudo-outcome of CATE, we first use Super Learner \citep{van2007super}  to estimate the conditional mean function of the outcome \(\bQ_{0}\) and the propensity score function $g_0$ by \(\bQ_{n}\) and $g_n$, respectively. The pseudo-outcome of CATE is given by 
\[
\eta(\bQ_n, g_n)(O) 
:= \frac{2A - 1}{g_n(A \mid W)} \big[Y - \bQ_n(A, W)\big] 
+ \bQ_n(1, W) - \bQ_n(0, W).
\]
This pseudo-outcome is doubly robust in the sense that $E[\eta(\bQ_n, g_n)(O)|W] = B_0(W)$ if either \(\bQ_n=\bQ_0\) or \(g_n=g_0\).
Next, we estimate the CATE function by regressing the pseudo-outcomes 
of the observed data, \(\eta(\bQ_n, g_n)(O_i)\), on their baseline covariates \(W_i\) 
using first-order HAL estimators, based on various versions of model selectors and pointwise estimation methods introduced in previous sections for estimating conditional mean functions. 
This yields an estimated CATE function \(B_n : [0,1]^d \to \R\), and a standard error estimator \(\tau_n : [0,1]^d \to [0,\infty)\).
For any new covariate \(\tilde{w} \in [0,1]^d\),
let $B_n(\tilde{w})$ denote the estimated CATE and $\tau_n(\tilde{w})$ its estimated standard error using the delta method. These estimates are then used to construct 95\% Wald-type confidence intervals, as is done for inference on conditional mean functions.

We ran 500 Monte Carlo simulations to evaluate the performance of the proposed methods with the following data generating distribution. The baseline covariates \(W = (W_1, W_2)\) are independently drawn from a uniform distribution in \([0, 1]\).  The true propensity score $g_0$ is defined as 
$
g_0 = [1 + F_{\text{Beta}}(W_1; 2, 4)]/4,
$
where \(F_{\text{Beta}}\) is the cumulative distribution function of the Beta distribution. 
The assignment of treatment \(A\) given $W$ follows \(\text{Bernoulli}(g_0(W))\).
The conditional mean functions of the outcome given the treatment and baseline covariates are
\[
E[Y \mid A = 1, W] = \zeta(W_1) \zeta(W_2),
\quad
E[Y \mid A = 0, W] = 0.9 \cdot \zeta(W_1) \zeta(W_2),
\]
where \(\zeta(w) =   1/(1+\exp(-12(w - 0.5)))\). 
The observed outcome is generated by
\[
Y = A \cdot E[Y \mid A=1, W] + (1 - A) \cdot E[Y \mid A=0, W] + \epsilon,
\quad
\epsilon \sim N(0, \sigma(W)^2),
\]
where $\sigma(W) = \sqrt{-\log(W_1)}$. We evaluate the performance at 20 different test points across sample sizes $n = 250, 500, 1000$ for fitting HALs.

\begin{figure}[tbh] 
    \centering
\includegraphics[width=1\linewidth]{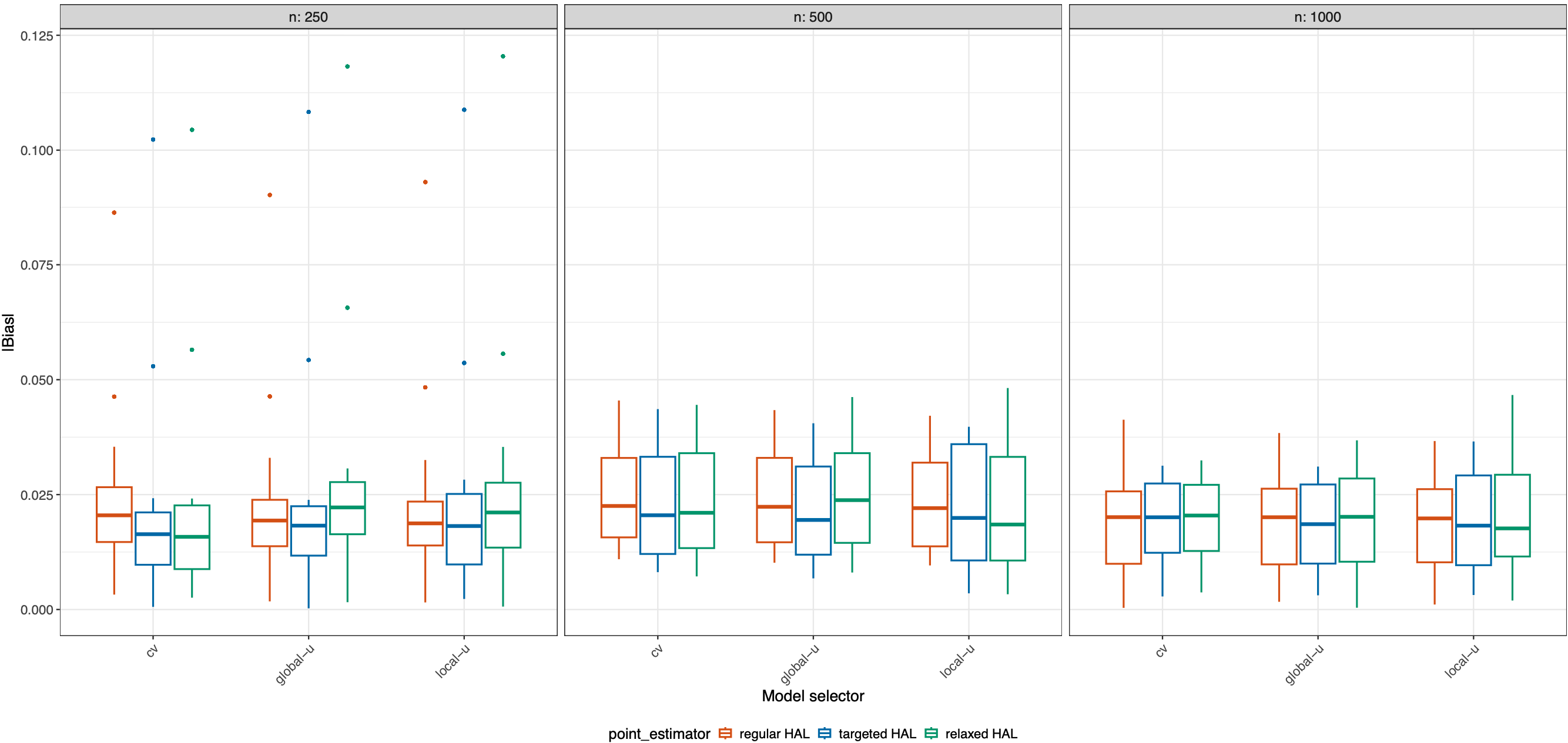}
    \caption{Bias of HAL-based point estimators for CATE across sample sizes (grids). Each grid displays box plots of bias (y-axis) of estimators by model selector (x-axis) and estimation method (color) across test points.}
    \label{fig:cate-bias}
\end{figure}

\begin{figure}[tbh]
    \centering
\includegraphics[width=1\linewidth]{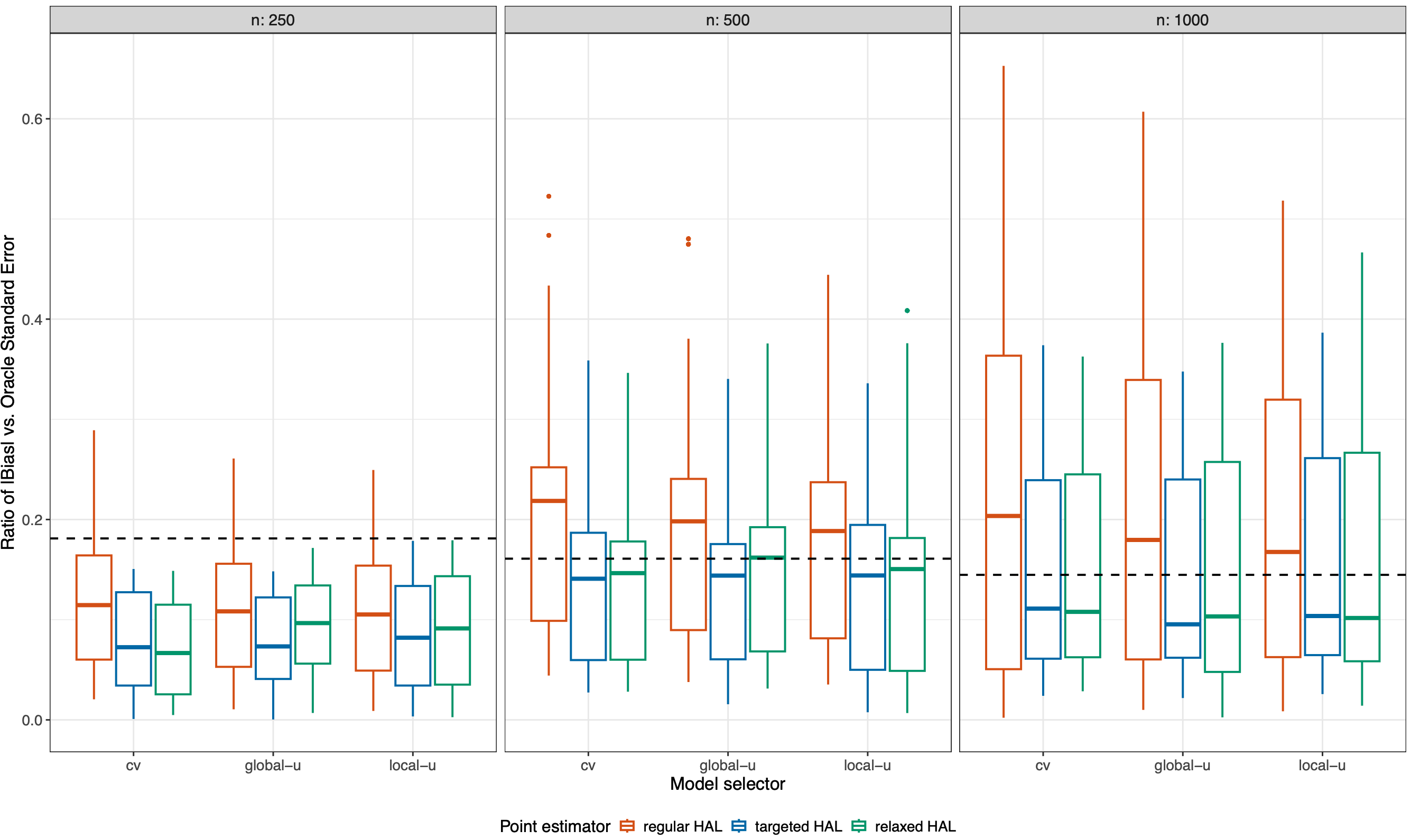}
    \caption{Ratio of bias to oracle standard error for HAL-based CATE estimators across sample sizes (grids). Each grid shows box plots of this ratio (y-axis) by model selector (x-axis) and estimation method (color) across test points.}
    \label{fig:cate-bias-se-oracle}
\end{figure}

\begin{figure}[tbh]
    \centering
\includegraphics[width=1\linewidth]{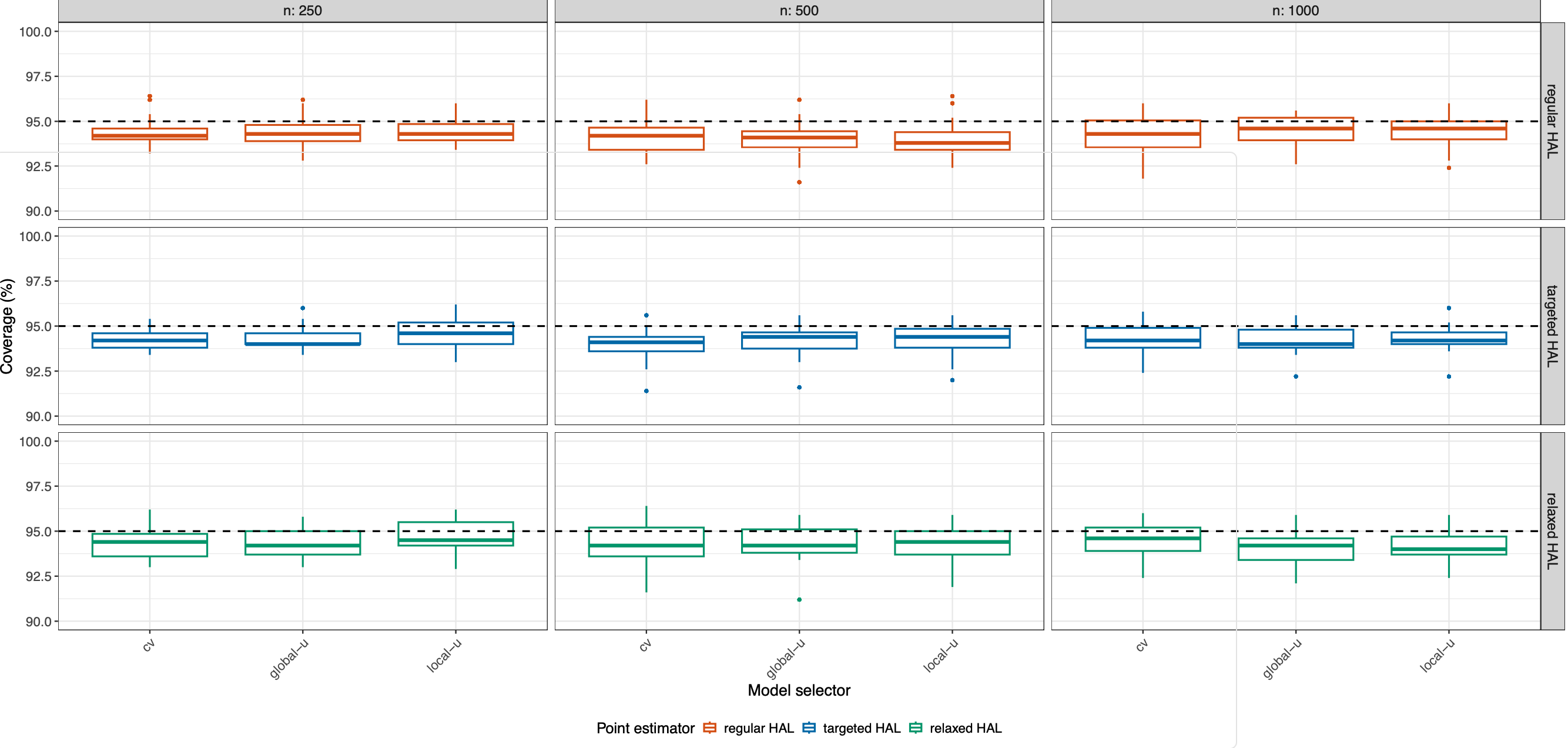}
    \caption{Coverage probability of oracle confidence intervals for HAL-based CATE estimators.
    Each panel shows a specific sample size (column) and estimation method (row),
    and displays box plots of the coverage probabilities of oracle confidence intervals (y-axis) by model selector (x-axis) across test points.}
    \label{fig:cate-cov-oracle}
\end{figure}

Figures \ref{fig:cate-bias}, \ref{fig:cate-bias-se-oracle}
and \ref{fig:cate-cov-oracle} present the bias, the ratio of bias versus oracle standard error and the coverage of the 95\% Wald-type confidence interval with oracle standard error. All methods show nominal coverage rate for their oracle confidence intervals, with undersmoothing and debiasing further contributing to bias reduction compared with standard error in most scenarios.

\begin{figure}[tbh]
    \centering
\includegraphics[width=1\linewidth]{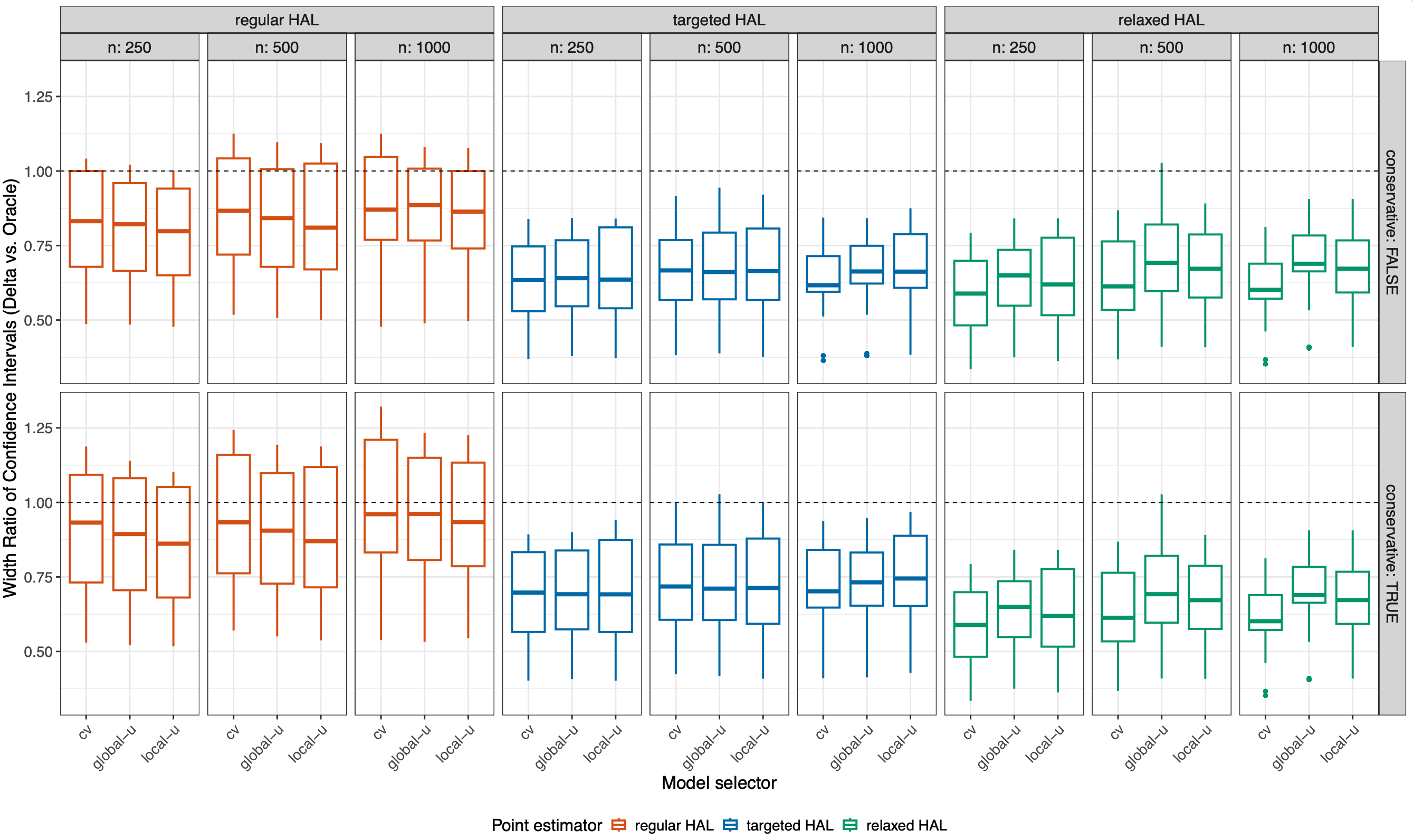}
    \caption{
    Width ratios of delta-method-based confidence intervals to oracle confidence intervals for CATE across estimation methods (columns), with each row representing a type of delta-method-based interval. Each sub-panel presents box plots of width ratios (y-axis) by model selector (x-axis) across sample sizes.}
    \label{fig:cate-width-ratio}
\end{figure}

\begin{figure}[tbh]
    \centering
\includegraphics[width=1\linewidth]{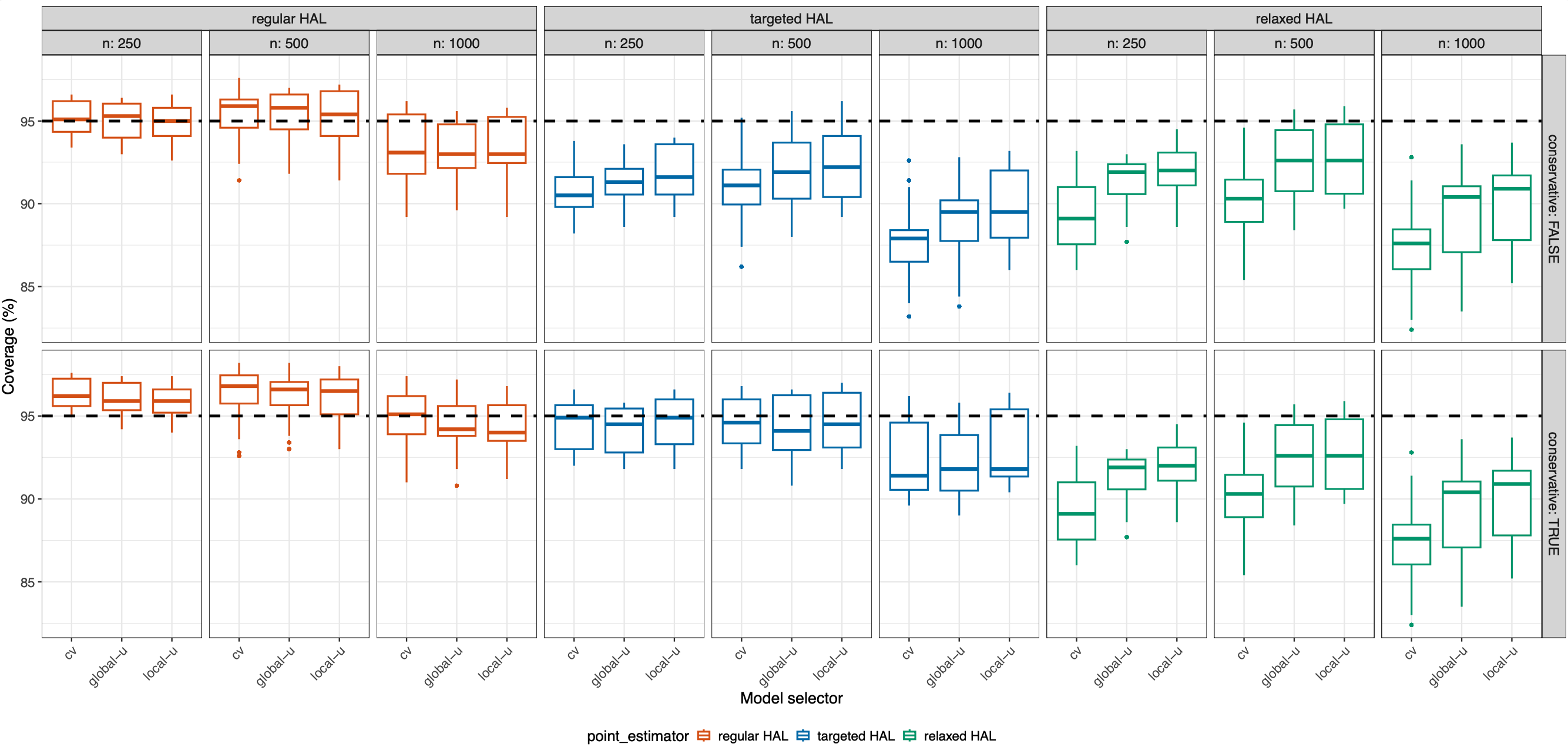}
    \caption{Coverage probability of delta-method-based confidence intervals for CATE across estimation methods (columns), with each row representing a type of delta-method-based interval. Each sub-panel presents box plots of coverage probabilities (y-axis) by model selector (x-axis) with different sample sizes.}
    \label{fig:cate-cov-delta}
\end{figure}

Figure \ref{fig:cate-width-ratio} presents the ratio of width of delta-method-based confidence intervals and oracle confidence intervals. It shows that the width of the delta-method-based confidence interval tends to be smaller than that of the oracle confidence interval for targeted HAL and relaxed HAL, while such width ratios for regular HAL are closer to 1. Figure \ref{fig:cate-cov-delta} displays the coverage probabilities delta-method-based confidence intervals. It shows that delta-method-based confidence interval for regular HAL achieves higher coverage rate than the targeted HAL and relaxed HAL, and stays around nominal coverage. The relative performance of their coverage probabilities aligns with what we find for conditional mean functions.

\section{Conclusion}
\label{sec:discussion}
In this study, we discuss different versions of HAL-based point estimators and methods to provide inference using the delta method.
For point estimators, we discuss regular HAL-MLE, targeted HAL, and relaxed HAL, and propose new model selection strategies, including global undersmoothing and local undersmoothing, tailored to reduce bias and support delta-method-based inference for estimating conditional mean functions based on HAL working models.

In the simulation study, estimators based on working models chosen by local undersmoothing showed the least bias, followed by those using global undersmoothing and then cross-validation.
Both targeted HAL and relaxed HAL significantly reduced bias compared to regular HAL under \(L_1\)-norm constraints. 
Applying undersmoothing techniques combined with targeted or relaxed HAL is the most effective strategy to reduce bias in most scenarios. These estimators also show the nominal coverage rate of their oracle confidence intervals. 

Across different combinations of model selectors and point estimators, we find that delta-method-based confidence intervals for regular HAL based on undersmoothed models showed the highest coverage probabilities. 
This performance results from effective bias reduction compared with standard error via undersmoothing together with the delta method's conservative variance estimation for the regular HAL.

Overall, this work contributes to the development of flexible point estimation and robust inference methods for conditional mean functions and other non-pathwise differentiable parameters by  leveraging Highly Adaptive Lasso, undersmoothing and debiasing strategies, as well as delta-method-based inference.
In summary, if the objective is to obtain a good point estimator with minimal bias of conditional mean (or other functional parameters of interest) based on HAL, we recommend using targeted HAL or relaxed HAL with the proposed undersmoothing techniques. 
If the goal is to make statistical inference of conditional means, we recommend constructing delta-method-based confidence intervals by regular HAL-MLEs under undersmoothed models. 
We also extend the use of the proposed methods for inference on other non-pathwise differentiable parameters by estimating the conditional average treatment effect functions. This lays the foundation for further investigation into robust statistical estimation and inference for various types of non-pathwise differentiable parameters using the Highly Adaptive Lasso.

\bibliography{ref.bib}

\end{document}

%% file: figures/bias.tex
\begin{figure}[tbh]
    \centering
    \includegraphics[width=1\linewidth]{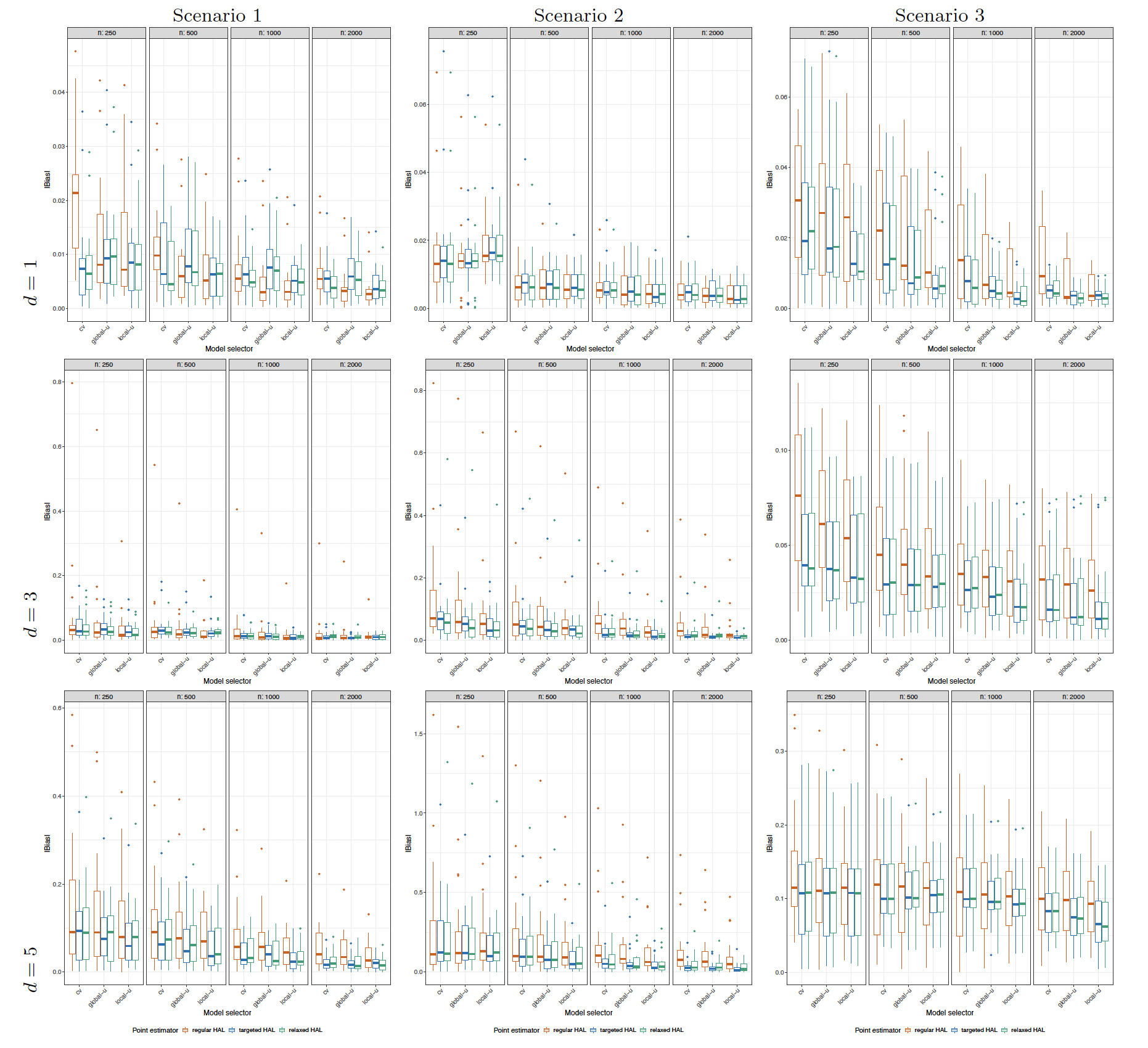}
    \caption{Bias of HAL-based point estimators under all scenarios. Each panel shows a specific data-generating distribution (column) with a given dimension (row). Within each panel, sub-panels represent different sample sizes. In each sub-panel, each box plot represents a different point estimator indexed by a model selector (x-axis) and estimation method (color) and shows its bias of that estimator across test points (y-axis).}.
    \label{fig:boxplot_bias}
\end{figure}

%% file: figures/bias_reg_proj.tex
\begin{figure}[tbh]
    \centering
    \includegraphics[width=1\linewidth]{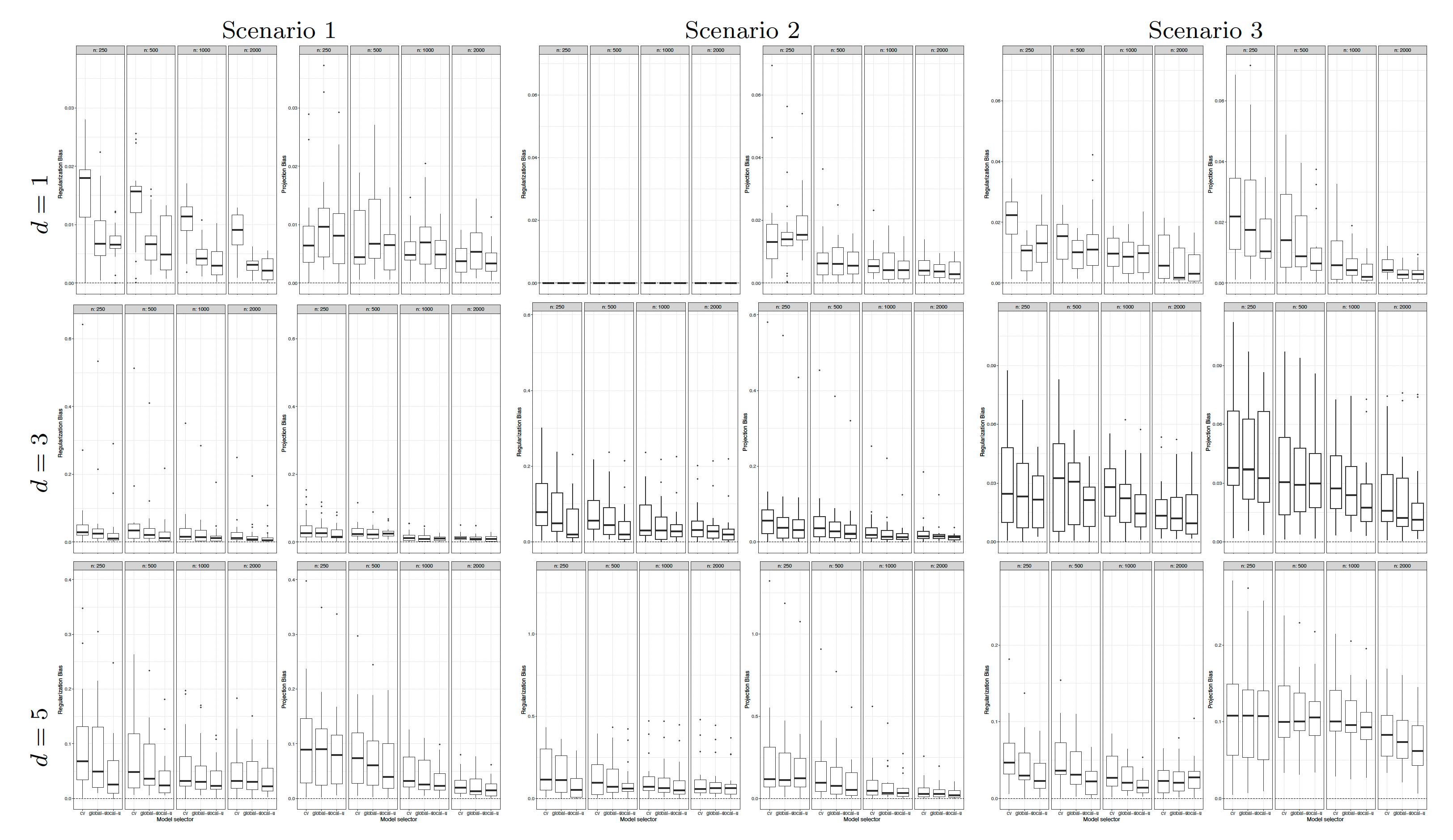}
    \caption{Regularization and projection bias of regular HAL estimators under all scenarios. Each panel shows a specific data-generating distribution (column) and dimension (row). Within each panel, the left sub-panels display box plots of regularization bias (y-axis) acros test points with different sample sizes, and the right sub-panels show box plots of projection bias (y-axis) across sample sizes, grouped by model selector (x-axis).}
    \label{fig:bias_reg_proj}
\end{figure}

%% file: figures/bias_se_oracle.tex
\begin{figure}[tbh]
    \centering
    \includegraphics[width=1\linewidth]{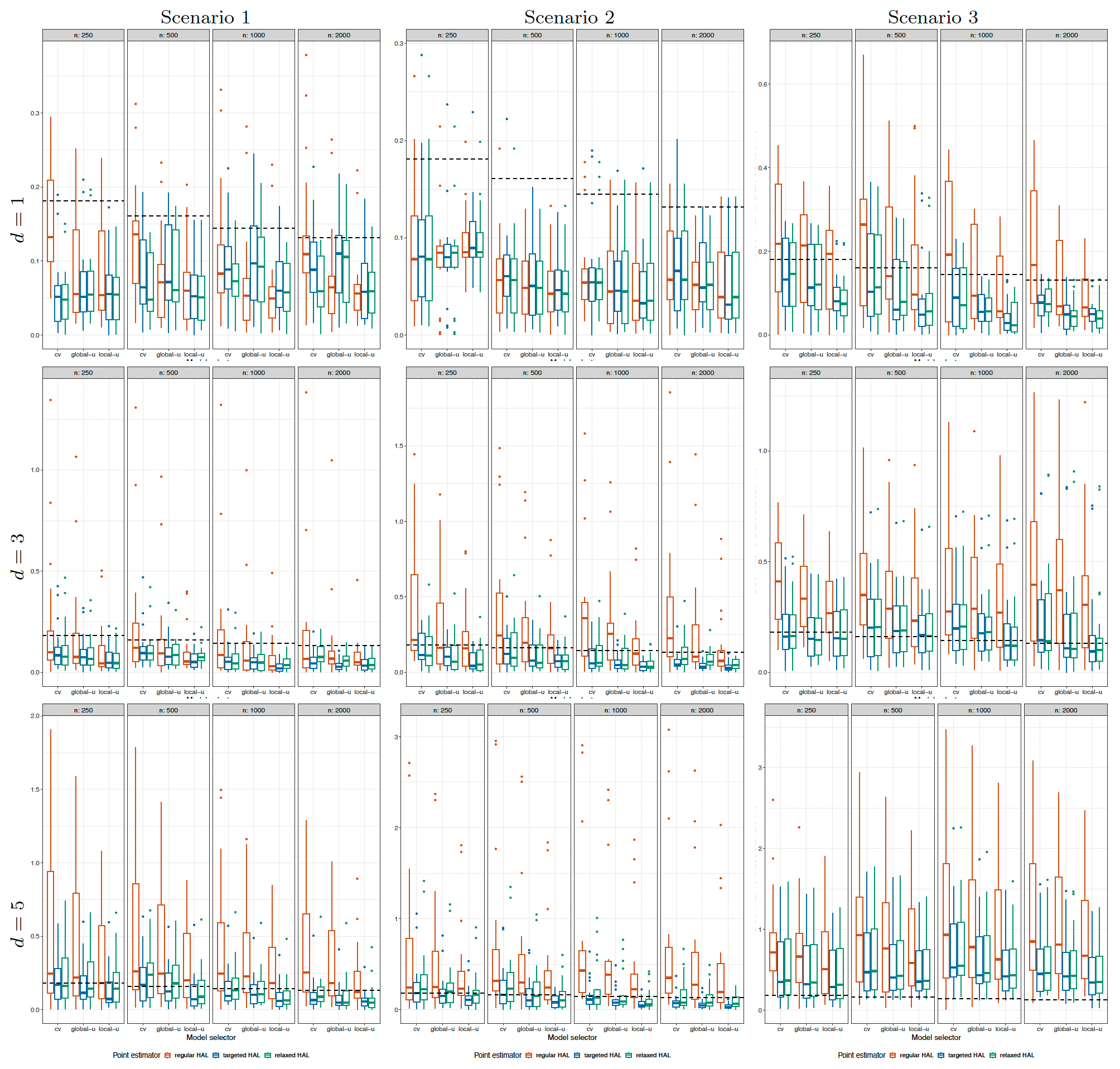}
    \caption{Ratio of Bias over Monte Carlo standard deviation of point estimators across 20 test points in all scenarios. 
    Each panel shows a specific data-generating distribution (column) with a given dimension (row). Within each panel, sub-panels represent different sample sizes. In each sub-panel, each box plot represents a different point estimator indexed by a model selector (x-axis) and estimation method (color) and shows its ratio of bias over Monte Carlo standard deviation of that estimator (y-axis).
    The dashed line ($1/\log(n)$) is the threshold used in the undersmoothing methods.}
    \label{fig:boxplot_bias_se_oracle}
\end{figure}

%% file: figures/cov_oracle.tex
\begin{figure}[tbh]
    \centering
    \includegraphics[width=1\linewidth,trim={0 1cm 0 0},clip]{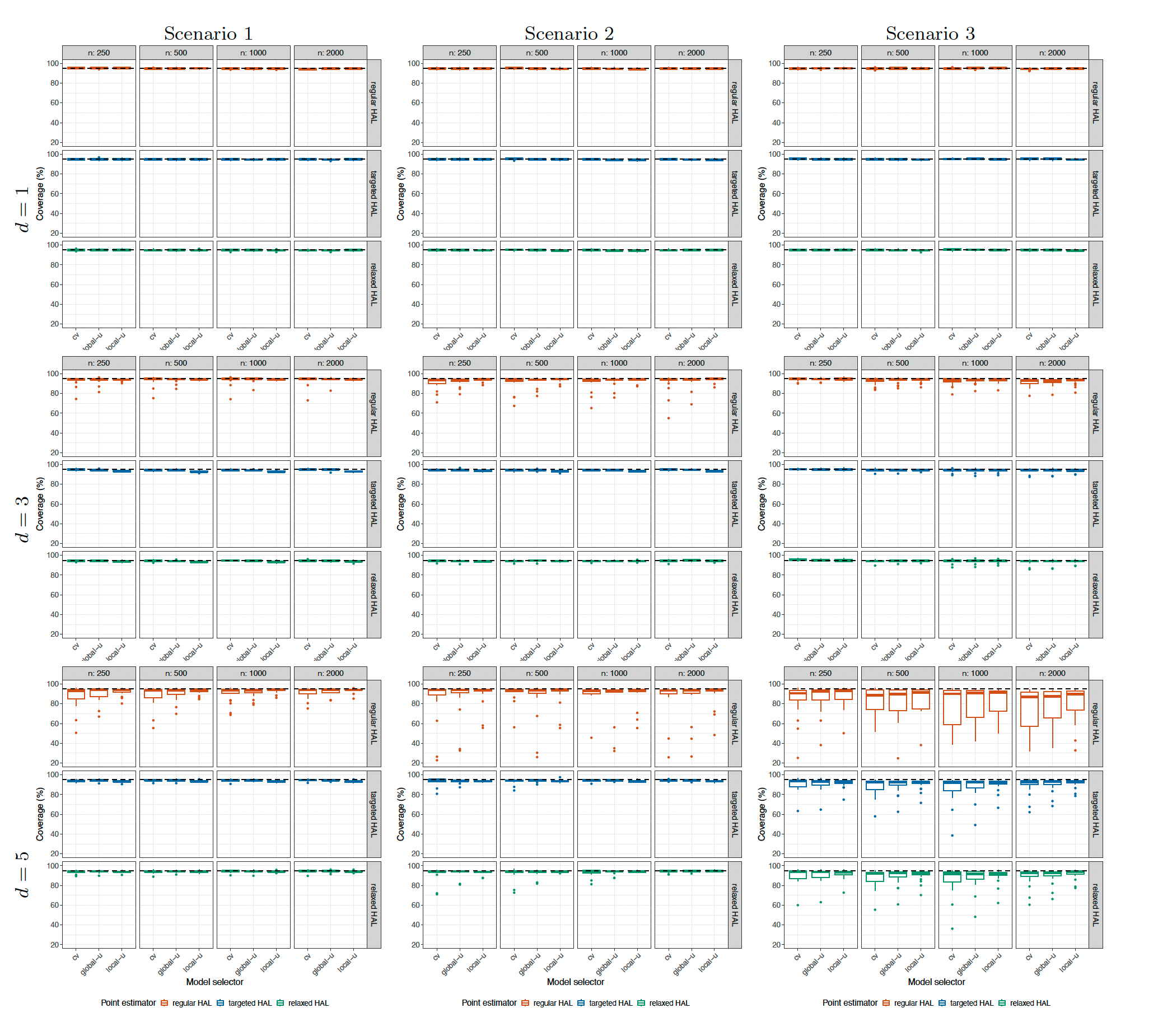}
    \caption{Coverage probability of oracle confidence intervals in all scenarios.
    Each panel shows a specific data-generating distribution (column) and dimension (row). Within each panel, sub-panels are arranged by estimation method (rows) and sample size (columns). Inside each sub-panel, the x-axis indicates the model selector, and the y-axis shows the coverage probability of oracle confidence intervals across test points.}
    \label{fig:cov_oracle}
\end{figure}

%% file: figures/width_delta_oracle.tex
\begin{figure}[tbh]
    \centering
    \includegraphics[width=1\linewidth]{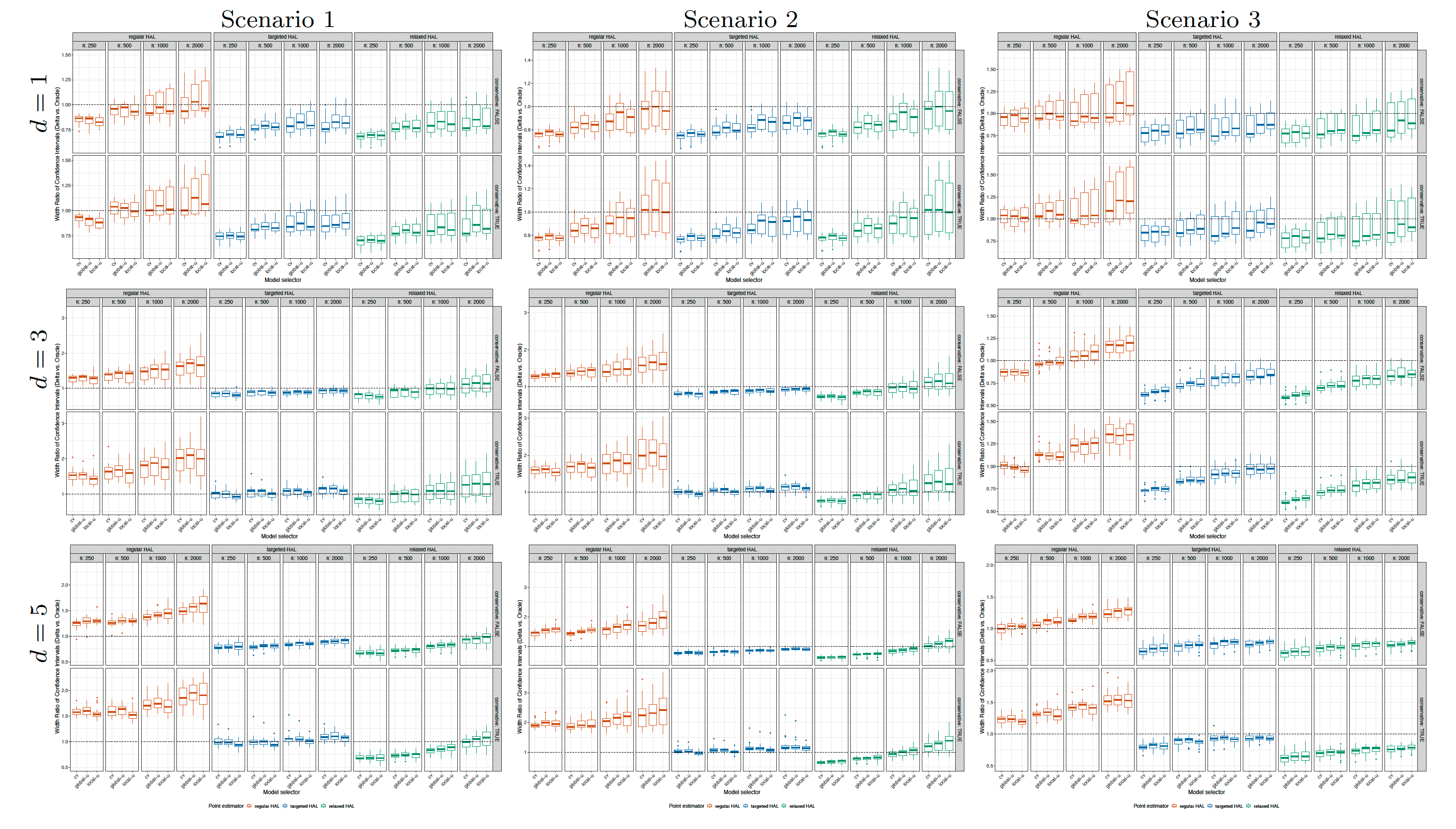}
    \caption{Width ratio of delta-method-based confidence intervals to oracle confidence intervals across all scenarios, dimensions, sample sizes, and point estimators. Columns indicate scenarios, rows indicate dimensions. Each sub-panel has six grids (column: estimation methods, row: type of confidence interval), showing box plots of width ratios (y-axis) by model selector (x-axis) across different sample sizes.}
    \label{fig:width_ratio}
\end{figure}

%% file: figures/cov_delta.tex
\begin{figure}[tbh]
    \centering
    \includegraphics[width=1\linewidth]{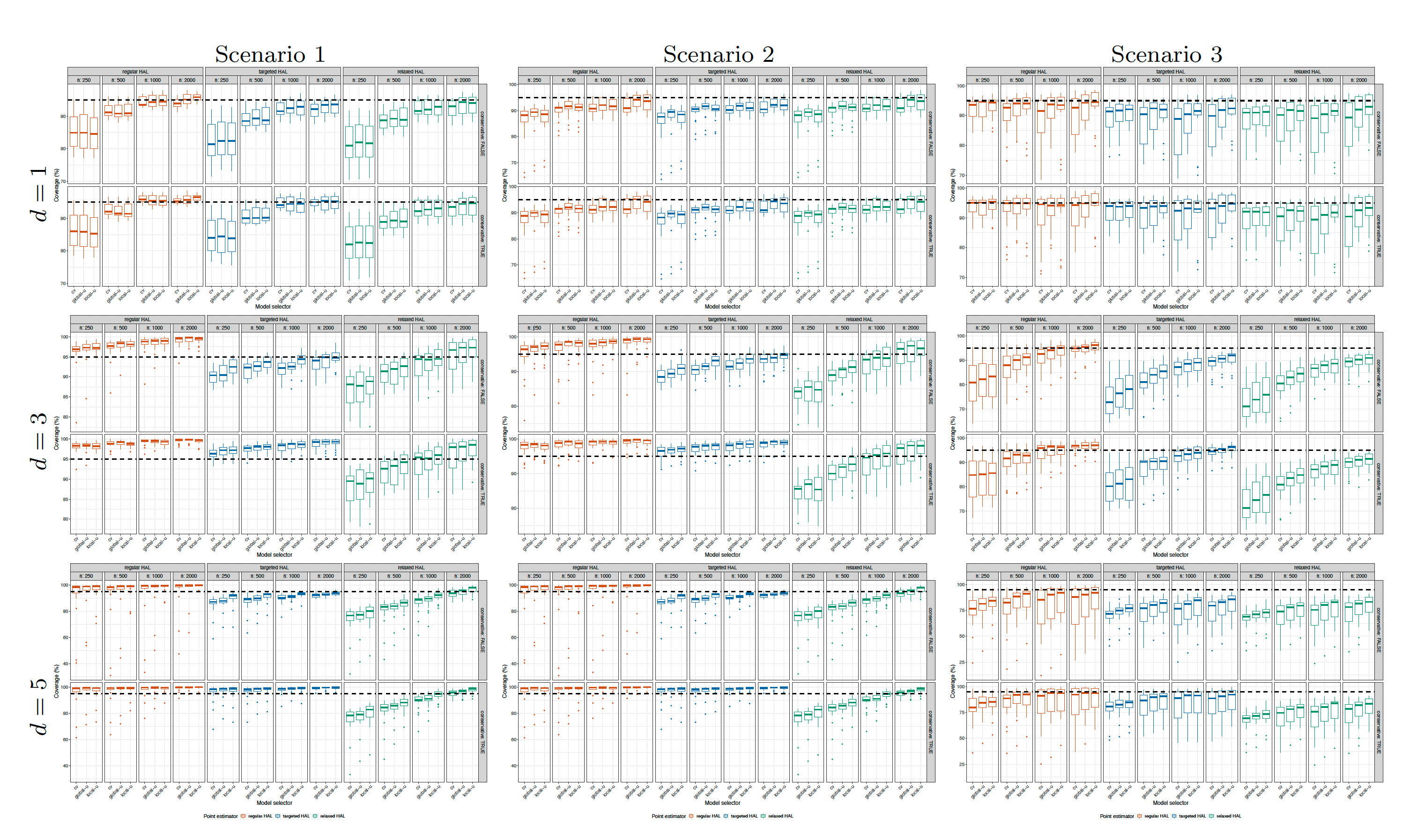}
    \caption{Coverage probability of delta-method-based confidence intervals across all scenarios, dimensions, sample sizes, and point estimators. Columns indicate scenarios, rows indicate dimensions. Each sub-panel has six grids (column: estimation methods, row: type of confidence interval), showing box plots of coverage probabilities (y-axis) by model selector (x-axis) across different sample sizes.}
    \label{fig:cov_delta}
\end{figure}